
\documentstyle{amsppt}
\magnification 1200
\def\pd#1#2{ \dfrac{\partial #1}{\partial #2} }
\def\tr{{\text{tr}}\,}
\def\tq{{\tilde q}}
\def\L{{\it L}}
\def\tl{{\tilde L}}

\def\TT{{\tilde T}}
\def\th{{\hat T}}

\def\LL{{\text \tensy L }}
\def\M{{\text\tensy M}}
\def\F{{\text\tensy F}}
\def\TL{{\tilde{\text{\tensy L}}}}
\def\TM{{\tilde{\text{\tensy M}}}}
\def\TF{{\tilde{\text{\tensy F}}}}
\def\overnumber#1#2{\mathop{\vbox{\ialign{##\crcr\noalign{\kern2pt}
      $\scriptstyle{#2}$\crcr\noalign{\kern2pt\nointerlineskip}\kern-2pt
      $\hfil\displaystyle{#1}\hfil$\crcr}}}\limits}

\def\overtimes{\mathop{\vbox{\ialign{##\crcr\noalign{\kern1pt}
      $\displaystyle{\otimes}$\crcr\noalign{\kern1pt\nointerlineskip}\kern-1pt
      $\hfil\displaystyle{,}\hfil$\crcr}}}\limits}
\def\qdet{\text{det$_q$}\,}

\document
Mathematical preprint series, Report 94-07, University of Amsterdam \par
February 7, 1994; \quad hep-th/9402111.
\vskip 10mm
{\bf Separation of variables for the
quantum relativistic Toda lattices}
\vskip 10mm
{V.B.~Kuznetsov${}^a$ and A.V.~Tsiganov${}^b$}
\vskip 5mm
${}^a$ {\it Department of Mathematics and Computer Science,
            University of Amsterdam, Plantage Muidergracht 24,
            1018 TV Amsterdam, The Netherlands;}\quad
            e-mail: {\text{vadim\@fwi.uva.nl}}
\vskip 5mm
${}^b$ {\it Department of Computational Physics,
            Institute of Physics, St.Petersburg University,
            St.Petersburg 198904, Russia;}\quad
            e-mail: {\text{tsigan\@onti.phys.lgu.spb.su}}
\vskip 10mm
Short title: Quantum relativistic Toda lattices
\vskip 10mm
{\bf Key words:} Toda lattices, separation of variables,
quadratic $R$-matrix algebras, reflection equation,
quantum inverse scattering method
\vskip 5mm
{\bf AMS classification:} 81R50, 82B23, 17B37, 17B67
\vskip 10mm
{\bf Abstract}
\noindent
We consider quantum analogs of the relativistic Toda
lattices and give new $2\times 2$ $L$-operators for these models.
Making use of the variable separation the spectral problem
for the quantum integrals of motion is reduced to solving
one-dimensional separation equations.

\head 1. Introduction
\endhead
\noindent
The relativistic Toda lattices (RTL's) were originally  introduced
in [9] where the classical nonperiodic case has been solved by means
of an explicit action-angle transformation with the help of the
$N\times N$ matrix Lax representation for the model.
For this RTL, which is naturally associated to the root system
$A_{N-1}$, the   Lax
representation in terms of $N\times  N$   matrices appeared to
have more
complicated structure in comparison  with the Lax pair for
corresponding non-relativistic
Toda lattice [8].  In  [1,2] there was introduced a  Lax triad
for the model, three matrices of which have simplier three-diagonal
form.  With the help of this   representation  the  direct   and
inverse problems for the system have been solved  and the
separation of variables has been carried  out in full analogy to   the
classical work [4].

In recent works [12-14] there was   proved  an equivalence,  due   to
canonical  transformation, of the RTL's for general
classical root systems
and discrete time
Toda lattices.  There was given also an   orbit
interpretation  of  the  Lax representations for all the systems.
Following [3,7,11]
in work [13]  for all the RTL's there    were
introduced  $2\times2$ $L$-operators
which satisfy the Sklyanin quadratic algebra  with non-standard
trigonometric $r$-matrix.

In the  present paper  we
construct, for all the RTL's, new $2\times 2$ quantum $L$-operators
obeying the qudratic $R$-matrix relations with the
standard trigonometric $R$-matrix.
We give also the separation  of variables for the $A_{N-1}$
RTL which is a generalization  to  relativistic  case  of  the
variable  separation  procedure  given  in  [10]  for non-relativistic
Toda lattice.

The structure of the  paper  is as  follows: in the Section 2
we recall the known facts about the classical RTL, which will be  needed
in the  sequel.   In  the Section 3 there  is  introduced
new $2\times   2$   $L$-operator  for the $A_{N-1}$ RTL, which is
a nonsymmetric $L$-operator of the lattice sh-Gordon model,  and
there is given also
the  variable  separation   procedure   indicating   that our
separation  variables  are  equivalent  to  the  ones introduced
before [1].    In the Section 4  we consider the quantum $A_{N-1}$
RTL and construct   the  separation  variables  and  corresponding
separation equations for it. The final Section 5
describes the construction of the monodromy matrix
for the  quantum  RTL's for general classical root systems
(including affine cases).

\head 2. Description of the model
\endhead
\noindent
The relativistic generalizations of the classical
Galilei-invariant Calogero-Moser systems and Toda lattices are
characterized by the time and space translation generators
$$\aligned
 T\equiv\frac12\,(S_+\,&+S_-\,)\,,\qquad
P\equiv\frac12\,(S_+\,-S_-\,)\,,\\
 \text{where}\qquad S_\pm
\,&\equiv \sum\limits_{j=1}^N\,
e^{\pm\theta_j}\,V_j\,(\tq_1,\ldots,\tq_N\,)\,,
\endaligned
\tag1$$
and the boost generator
$$B\equiv - \sum\limits_{j=1}^N\,\tq_j\,.
 \tag2$$
In this equations
$\theta_j$  and  $\tq_j$ are canonically conjugated
momentum and coordinate of the $j$-th particle.
One obtains a representation of the Lie algebra of the Poincare
group if and only if $V_j$ satisfy the functional
equations [9]
$$\aligned &V_i\,\partial_i\,V_j\,+
V_j\,\partial_j\,V_i\,=0\,,\qquad i\neq j\,, \\
&\sum\limits_{j=1}^N\,\partial_j\,V^2_j\,=0\,,\qquad
\partial_j=\pd{}{\tq_j}\,.
\endaligned
 \tag3$$

There are known two solutions to this equation [9]:
$$\align
&\text{1.}~V_j\,(\tq_1\,,\ldots,\tq_N\,)=
\prod\limits_{i\neq j}\,f\,(\tq_j-\tq_i\,)\,,~
f^2\,\equiv a+b\wp\,(\tq)\tag4 \\
&\text{where}~\wp~ \text{ is the Weierstrass function.}\qquad \\
&\text{2.}~V_j\,(\tq_1\,,\ldots,\tq_N\,)=
f\,(\tq_{j-1}\,-\tq_j)\,f\,(\tq_j\,-\tq_{j+1}\,)\,, \tag5 \\
&f\,\equiv (1+g^2\,e^{\tq}\,)^{1/2}\,,\qquad g\in R\,. \\
\endalign
$$
The first case is called relativistic Calogero-Moser system
and the second case is referred to as RTL.

As phase space we have
$$\Omega\equiv\{(\theta\,,\tq)\,\in R^{2N}\,\}\,,\qquad
\omega\equiv\sum\limits_{j=1}^N\,d\theta_j\,\wedge\,d\tq_j\,.
\tag6$$

We will consider RTL as the integrable system of $N$
one-dimentional particles with the Hamiltonian
$$H =\sum_{j=1}^M\,\,\exp\,(\theta_j\,)
\Bigl[\bigl[1 + \exp(\tq_j\,-\tq_{j-1}\,)\,]\,
[\,1+\exp(\tq_{j+1}\,-\tq_j\,)\,\bigr]\Bigr]^{1/2}\,.
\tag7$$
The RTL is open (non-periodic)
when $M=N-1$  and periodic in the case of
$$M=N,~\tq_{j+kN}\,=\tq_j\,,~\theta_{j+kN} = \theta_j,~\text{ where}~
j = 1,\ldots,N\,~\text{ and}~k\in \text{{\bf Z}}\,.
\tag8$$
In this section we consider only periodic RTL.

According to [12,13] there is an equivalence between the RTL
and the discrete time Toda lattice.
It appeared to be more
natural to introduce new canonical coordinates ($p_j\,,\tq_j$)
connected to the old coordinates ($\theta_j\,,\,\tq_j$) as follows:
$$p_j\,=\theta_j\,+\frac12\,\ln\left(
{\frac{1+\exp(\tq_j-\tq_{j-1}\,)}{1+\exp(\tq_{j+1}\,-\tq_j\,)}}\right)\,.
\tag9$$
Transformation $(\theta_j\,,\,\tq_j\,)\rightarrow
(p_j\,,\tq_j\,)$ is a canonical transformation.

In new variables the Hamiltonian (7) has the form
$$H=\sum_{j=1}^N\,\exp(p_j\,)\Bigl[1+\exp(\tq_{j+1}\,-\tq_j\,)\,\Bigr]\,.
\tag10$$
One can introduce a set of new variables $c_j\,,~d_j$ and $f_j$ [1,2]
$$c_j = \exp(p_j-\tq_j+\tq_{j+1})\,,\qquad  d_j = \exp p_j\,,
\qquad f^2_j= c_j\,.
\tag11$$
Then the equations of motion given by the Hamiltonian (10)
have the following form
$$\aligned\dot{d_j}&= - d_j\, ( c_{j-1} - c_j)\,,\\
\dot{c_j}& = c_j\,( d_{j+1}\,- d_j\, + c_{j+1} - c_{j-1})\,,
\endaligned
\tag12$$
and can be rewritten as a compatibility condition for the two
linear problems:
$$\aligned
(k^2& - d_j)\,\phi_j + k\,( f_j\,\phi_{j+1} +
f_{j-1}\,\phi_{j-1}\,)= 0\,,\\
\dot{\phi}_j\,&= - {\frac12}\,(c_j\, - c_{j-1} + k^2)\,\,\phi_j -
{\frac12}\,k\,( f_j\,\phi_{j+1} - f_{j-1}\,\phi_{j-1})\,,
\endaligned
\tag13$$
where $\phi^t\,=(\phi_1\,,\ldots,\phi_N\,)^t$ is so-called
Baker-Akhiezer function.

Let us give the equations (13) in the matrix form introducing
two $N\times N$ matrices $\TL\,(k )$ and $\TM\,(k )$
$$\TL\,(k)\,\phi = 0\,,\qquad \dot\phi = - \TM\,(k)\,\phi\,.
\tag14$$
The equations of motion are thereby written in terms of the Lax
triad ($\TL,\,\TM,\,\TF$) [1,2]:
$$\dfrac{d~~}{dt}\,\TL =\Bigl[\,\TL\,,\,\TM\,\Bigr] + \TF\,\TL\,,
\tag15$$
where the symmetric matrix $\TL$ has the form
$$\TL=\pmatrix  {k^2-d_1}&{k\,f_1} &0       &\ldots      &{k\,f_N}\\
                 {k\,f_1}&{k^2-d_2}&{k\,f_2}&\ldots      &0       \\
                 \vdots  &\ddots   &\ddots  &\ddots      &{k\,f_{N-1}} \\
                 {k\,f_N}&0        &\ldots  &{k\,f_{N-1}}&{k^2-d_N}
				 \endpmatrix\,,
\tag16$$
and the matrix $\TM$ looks like
$$\TM=\dfrac12
\pmatrix  c_1-c_N+k^2&{k\,f_1} &0             &{-k\,f_N}\\
               {-k\,f_1}&{c_2-c_1+k^2}&{k\,f_2}  &0       \\
               \ddots  &\ddots  &\ddots          &{k\,f_{N-1}} \\
               {k\,f_N}&0        &{-k\,f_{N-1}}   &{c_N-c_{N-1}+k^2}
\endpmatrix \,.$$
The matrix $\TF$ is diagonal and has zero trace:
$$\TF=\pmatrix  {f_1^2-f_N^2}&~ &~      \\
                 ~      &{f_2^2-f^2_1}&~      \\
                 ~      &~\ddots         &~      \\
                 ~      &~         &{f^2_N-f^2_{N-1}}
\endpmatrix
				 \,.$$

{}From the Lax representation (15) there follows that
$$\tr\,\TL^{-1}\,\dot\TL = \tr\, \TF = 0\,,
\tag17$$
and, therefore, using the Abel identity
$${\pd{\ln\det X}{t}} =\tr\, X^{-1}\,{\pd{X}{t}}\,,
\tag18$$
one concludes that the $\det\,(\TL)$ is the generating function
of the integrals of motion for the RTL.  The Lax triad (15) and
the related questions were studied in [1,2]. We are giving below
one Proposition from there.

\proclaim{Proposition 1}
\flushpar
 a).  The  determinant  and  all  principal minors of the matrix
 $\TL   \,(k   )$   are   polynomials   on   $k^2$.
\flushpar
  b). The zeros of determinant and of all principal minors of
 the matrix $\TL \,(k )$ are simple and positive real.
\endproclaim

According to [3,7] it is possible to rewrite the equations (13)
in the matrix form with the help of the $2\times 2$
matrices $\tl_j\,(k )$ [12]
$$\tl_j\,(k)=\pmatrix k^2-\exp(p_j)\,,&-k\,\exp(\tq_j\,)\\
                    k\,\exp(p_j-\tq_j\,)\,,&0
\endpmatrix\,.
\tag19$$
The matrix $\TL_N\,(k)$ is related to
the $2\times 2$ monodromy matrix $\TT_N\,(k )$ such that
$$\det\,\TL_N\,( k )-2 = \tr\,\TT_N\,( k )\equiv
\tr\,\prod^N_{j=1}\tl_j\,( k )\,=\tr
\pmatrix \tilde{A}_N&\,\tilde{B}_N\,\\\,\tilde{C}_N&\,\tilde{D}_N
\endpmatrix\,.
\tag20$$
Matrices $\tl_j\,(\lambda)$  and monodromy  matrix $\TT_N  \,(\lambda)$
satisfy the Sklyanin quadratic algebra [3,10]
$$\Bigl\{\TT^{(1)}\,(\lambda)\,,\TT^{(2)}\,(\mu)\Bigr\}=
\Bigl[\tilde{r}\,(\lambda/\mu)\,,\TT^{(1)}\,(\lambda)\,\TT^{(2)}\,(\mu)\Bigr]
\tag21$$
where $\TT^{(1)}\,(\lambda)=\TT(\lambda)\,\otimes\,I$ and $\TT^{(2)}\,(\mu)=
I\,\otimes\,\TT(\mu)\,$,
with the following non-standard $r$-matrix
$$\tilde{r}(\lambda/\mu)=\pmatrix a&0&0&0~\\ 0&-1/2&c&0~\\
0&c&1/2&0~\\ 0&0&0&{a}~\endpmatrix\,,
\tag22$$
where
$a\,(\lambda/\mu)={\frac12}\,{\frac{\lambda^2+\mu^2}{\lambda^2-\mu^2}}$,
$c\,(\lambda/\mu)={\frac{\lambda\,\mu}{\lambda^2-\mu^2}}$.
Note that the algebra (21) can be rewritten in the form
$$\Bigl\{\TT(\lambda)\,\overtimes\,\TT\,(\mu)\Bigr\}=
\Bigl[r\,(\lambda/\mu)\,,\TT\,(\lambda)\,\otimes\,\TT\,(\mu)\Bigr]+
\Bigl[s\,,\TT\,(\lambda)\,\otimes\,\TT\,(\mu)\Bigr]\,.
\tag23$$
In equation (23) $r(\lambda/\mu)$ is the standard $r$-matrix of the $XXZ$ type
and $s$ is the matrix of scalars:
$$   r(\lambda/\mu)=\pmatrix a&0&0&0~\\ 0&0&c&0~\\
0&c&0&0~\\ 0&0&0&a~\endpmatrix\,,\qquad
s=\pmatrix 0&0&0&0~\\ 0&-1/2&0&0~\\
0&0&1/2&0~\\ 0&0&0&0~\endpmatrix\,.\tag24$$
In the quantum case the equation  (23) goes to the
nonstandard quadratic relation
$$R_1\,(\lambda/\mu)\,{T}^{(1)}\,(\lambda)\,{T}^{(2)}\,(\mu)
={T}^{(2)}\,(\mu)\,{T}^{(1)}\,(\lambda)\,R_2\,(\lambda/\mu)\,.
\tag25$$

We can express the
entries of the  monodromy matrix $\TT\,(\lambda)$ in terms of the determinant
and principal  minors of   the matrix   $\TL\,(k)$ in the same way as it  was
done  in   [7]   for the non-relativistic Toda   lattice.  Then using the
Proposition 1 we have the following properties
for the entries of the $\TT(\lambda)$.
\proclaim{ Proposition 2}
\flushpar
 a) $\tilde{A}_N\,(\lambda)   =\,\lambda^N+\ldots$
 is a polynomial of degree $N$ on the
  $\lambda = k^2$ where:
    $$\tilde{A}_N\,(0)=(-1)^N\,\exp(P)\,,\quad P = \sum^N_{j=1}\,p_j\,.$$
\flushpar
 b) $\tilde{B}_N\,(\lambda) = k\,(\beta\lambda^{N-1}+\ldots\,),\,$
    $\tilde{C}_N\,(\lambda) = k\,(\gamma\lambda^{N-1}+\ldots\,),\,$
    $\tilde{D}_N\,(\lambda) = \delta\lambda^{N-1}+\ldots.$
\flushpar
 c)  Zeros of the polynomials $\tilde{A}(\lambda),\,\tilde{B}(\lambda)/k,\,
\tilde{C}(\lambda)/k,$ and $\tilde{D}(\lambda)$
are simple and positive real.
\endproclaim

These facts will be used in the next section for doing
the separation of variables.
We will give also a new
$2\times 2$ $L$-operator and a new monodromy matrix
for the RTL which are closely connected to the previous ones,
but satisfy the standard $XXZ$-type quadratic $r$-matrix
algebra.

\head 3. Classical mechanics\endhead
\noindent
It is well-known that the $2\times 2$ $L$-operator for the  non-relativistic
Toda  lattice   can  be   obtained  by   contraction  from   the
$L$-operator  of  the  $XXX$-type  Heisenberg  magnet.  For  the
$XXZ$-type  Heisenberg  magnet  one  can  use  the analogous
contraction and construct the $L$-operator of the form:
$$L_j\,(u)=\pmatrix  2\sinh(u-{\dfrac{p_j}2})\,&-\exp(q_j\,)\\
                    \exp(-q_j\,)\,&0\endpmatrix\,,
\tag26$$
where the $p_j$ and $q_j$ are the canonically conjugated momentum
and coordinate of $j$-th particle. This $L$-operator obeys the
fundamental Poisson brackets (21) with the standard $r$-matrix
$r(\lambda/\mu)$
of the $XXZ$-type (24) ($\lambda=e^u$)
and it can be thought of as a nonsymmetric analog of the
$L$-operator for the lattice sh-Gordon model [6].

The $L$-operator (26) and the monodromy matrix $T_N\,(u)$,
$$T_N\,(u)=\prod^N_{j=1}\,L_j\,(u)=
\pmatrix  \,{A}&\,{B}\,\\\,{C}&\,{D}\,\endpmatrix\,(u)\,,
\tag27$$
have a simple connection to the $\tl_j\,(k)$-operator (19)
and the monodromy matrix $\TT\,(k)$
(20) through the following change of variables:
$$ \aligned q_j&=\tq_j-{\dfrac{p_j}{2}}\,,\qquad k = \exp( u )\,;\\
L_j\,(u)&= k^{-1}\,\exp(-{\frac{p_j}{2}})\,\tl\,(u)\,,\\
T_N\,(u)&= k^{-N}\,\exp(-{\frac{P}{2}})\,\TT_N\,(u)\,,\endaligned
\tag28$$
where $P  =  \sum^N_{j=1}\,p_j$  is the total momentum of the system
which is an integral of motion.

We can associate the following $N\times N$
matrix $\LL\,(u)$ to the monodromy matrix $T_N\,( u )$ (cf. with [7]):
$$\LL=\pmatrix
{2\sinh(u-p_1/2)} &e_{21}            &0          &e_{1N}           \\
 e_{21}           &{2\sinh(u-p_2/2)} &e_{32}     &0                \\
 \vdots           &\ddots            &\ddots     &e_{NN-1}         \\
 e_{1N}           &0                 &e_{NN-1} &{2\sinh(u-p_N/2)}\endpmatrix
\,,$$
where $e_{jn}\,= \exp({\dfrac{q_j-q_n}2})$.
The matrices
$\LL\,( u )$ and  $\TL\,( k )$ are connected by the formula
$$\LL\,( u ) = U\,\TL\,( k )\,U
\tag29$$
where the matrix $U$ has the form
$$ U = k^{-1/2}\,\text{diag}\bigl( \exp({-\dfrac{p_1}4}),
\exp({-\dfrac{p_2}4}),\ldots, \exp({-\dfrac{p_N}4})\,\bigr)\,. $$
Due to the relation (29) one can construct the Lax triad for the matrix
$\LL\,(u)$:
$${\dfrac{d\LL\,(u)}{dt}}
=\Bigl[\LL\,,\M\Bigr]+\F\LL\,,$$  where the matrices $\M$  and   $\F$
are expressed in terms of the known matrices $\TM$ and $\TF$
(see (15)--(16))
$$\aligned
\M&=U^{-1}\TM U+U^{-1}\dot{U}\,,\\
\F&=U^{-1}\TM U - U\TM U^{-1}+\TF +2 U^{-1}\dot{U}\,.\endaligned
\tag30$$
Notice that in the limit to non-relativistic case
the Lax triad
$\LL\,,~\M\,,~\F$ turns into well-known Lax pair for
the usual Toda lattice [4,8].

The separation  of variables for the periodic RTL  was   originally
given in [1].   Let us carry out this procedure
within the $r$-matrix method with an intention to generalise all
the results afterwards to the quantum case.

We begin with the monodromy matrix $\TT\,(  k  )$ (20)--(21),
since due to the Proposition 2 the entries of it have simple
dependence upon the spectral parameter $k$. Let us take as new variables
$\mu_j$ the zeros of the off-diagonal entry
$\tilde   C\,(\lambda)$ (for such chosen, the variables $\mu_j$ coincide
up to a factor with the ones introduced earlier [2]
and are zeros of the Baker-Akhiezer function):
$$\tilde{C}\,(\mu_j) = 0,\qquad j = 1,\ldots,N-1\,.
\tag31$$
By the Proposition 2 these variables are positive real and
independent. Further, define a set of the variables $\nu^\pm_j$
as follows:
$$\nu^-_j\,= \tilde{A}_N\,(\mu_j\,)\,,
\quad\nu^+_j\, = \tilde{D}_N\,(\mu_j\,)\,,\quad j = 1,\ldots,N - 1\,.
\tag32$$
The variables $\mu_j\,,~\nu^\pm_j\,,~j  =  1,\ldots,N-1$
 obey the Poisson brackets:
$$\aligned
\bigl\{\mu_j\,,\,\nu^\pm_n\,\bigr\}&=\pm\delta_{jn}\,\nu^\pm_n\,,
\quad\bigl\{\nu^\pm_j\,,\,\nu^\pm_n\,\bigr\}=\bigl\{\mu_j\,,\,\mu_n\,\bigr\}
= 0\,,\\
\bigl\{\nu^-_j\,,\,\nu^+_n\,\bigr\}&=
-\delta_{jn}\,\left.{\dfrac {d\,\det\TT\,(\lambda)}{d\lambda}}
\right|_{\lambda
=\mu_j}\,,\\
\nu^{+}_j\,\nu^{-}_j&=\det\,\TT\,(\mu_j\,)\,.\endaligned
\tag33$$
These brackets follow from the fundamental Poisson brackets (21)
and can be calculated along the lines of the work [10].

Introduce two more variables
$$\mu_N\,= \exp ( p_N - \tq_N\,)\,,\quad\nu^\pm_N\,= \exp(\,\pm P).$$
The set of the $\mu_j\,,~\nu_j^\pm,~j = 1,\ldots,  N$ is
complete, i.e. any dynamical variable can be expressed through
them.  To prove this let us restore the monodromy matrix
$\TT\,( k  )$  in terms of the new variables:
$$\align
\tilde{A}_N\,(\lambda)&=\lambda\,\nu^-_N\,\prod^{N-1}_{\displaystyle j=0}\,
(\lambda-\mu_j)\,+\sum^{N}_{\displaystyle j=0}\,
\prod^{N-1}_{{\displaystyle i=0}\atop{\displaystyle i\neq j}}
\,\dfrac{\lambda-\mu_j}{\mu_i-\mu_j} \,\nu^-_j\,,\\
\intertext{ where $\mu_0\,\equiv 0$ and $\nu^-_0\,\equiv 1$,}
{}~&~\tag34\\
\tilde{D}_N\,(\lambda)&= \sum^{N-1}_{\displaystyle j=1}\,
\prod^{N-1}_{{\displaystyle i=1}\atop{\displaystyle i\neq j}}
\,\dfrac{\lambda-\mu_j}{\mu_i-\mu_j} \,\nu^+_j\,,\qquad
\tilde{C}_N\,(\lambda)=\mu_N\,k\,\prod^{N-1}_{\displaystyle j=0}\,
(\lambda-\mu_j)\,\endalign
$$
(recall that $\lambda\,=k^2$).  The entry $\tilde{B}\,(k)$
of the matrix $\TT\,(   k   )$ can be obtained using the identity
$\det\,\TT\,(k)=k^N\,\exp(\,-P)\,.$

Now we introduce another set of variables $y_j\,,~Y^\pm_j$ using the
monodromy matrix $T\,( u  )$ (26)--(27):
$$C\,(y_j\,)=0\,,\quad Y_j^-=A\,(y_j\,)\,,\quad Y_j^+ = D\,( y_j\,)\,,\quad
j=1,\ldots,N-1\,.
\tag35$$
The matrix entry $C\,(  u  )$, considered as a function on the
spectral parameter $u$, has infinite number of zeros, but using the
connection between the matrices $T\,( u )$ and  $\TT\,(  k  )$
(28) we can prove that these zeros have the following structure
$$ y_{jn}=\,y_j + 2\pi\,i\,n\,\equiv
 \ln \mu_j + 2\pi\,i\,n\,,\qquad n = 0, 1, 2,\ldots,
\tag36$$
hence the variables $y_j\,,~j = 1,\ldots,N - 1$
can be chosen to be simple and real.
Variables $y_j\,,~Y^\pm_j$ obey the Poisson brackets:
$$\bigl\{ y_j\,, y_n\,\bigr\} = \bigl\{ Y^\pm_j\,, Y^\pm_n\,\bigr\} = 0\,,
\quad Y^\pm_jY^\mp_j\,= 1\,,\quad\bigl\{y_j\,,Y^\pm_n\,\bigr\}=
\pm\delta_{jn}\,Y^\pm_n.$$
One can restore the monodromy matrix
$T\,(u)$ in terms of these variables.

The set $y_j\,,~Y^\pm_j$ gives the separation variables,
since by their definition (35), they obey the equations
$$ \tr\,T_N\,(y_j\,) = Y^+_j + Y^-_j\,,\qquad j = 1,\ldots,N-1\,.
\tag37$$

\head 4. Separation of variables in the quantum case
\endhead
\noindent
In this section we apply the general scheme [10] of the quantum
separation of variables to the quantum $A_{N-1}$ RTL model.

Quantum \L-operator for the non-relativistic Toda lattice [3,10]
has the form
$$L_j\,(u)=
\pmatrix u-{\dfrac{p_j}2}\,&-\exp(q_j\,)\\  \exp(-q_j\,)\,&0\endpmatrix \,.$$
For the RTL the quantum \L-operator and the monodromy matrix
$T_N\,( u )$ are
$$\aligned
L_j\,(u)&=\pmatrix 2\sinh(u-{\dfrac{p_j}2})\,&-\exp(q_j\,)\\
                    \exp(-q_j\,)\,&0\endpmatrix \,,\\
T_N\,(u)&=\prod^N_{j=1}\,L_j\,(u)=
\pmatrix \,{A}&\,{B}\,\\ \,{C}&\,{D}\,\endpmatrix \,, \,{\hbox{\rm where}}
\quad\bigl[\,p_j\,,q_n\,\bigr]=\,i\,\eta\,\delta_{jn}
\endaligned
\tag38$$
and obey the quadratic $R$-matrix algebra
$$R\,(u-v)\,T^{(1)}\,(u)\,T^{(2)}\,(v)
=T^{(2)}\,(v)\,T^{(1)}\,(u)\,R\,(u-v)\,,
\tag 39$$
where $T^{(1)}\,(u)=T(u)\,\otimes\,I$ and
$T^{(2)}\,(v)=I\,\otimes\,T(v)\,$ and $R(u)$
is the following trigonometric solution of the quantum
Yang-Baxter equation [3,6,10]:
$$R\,(u)=\pmatrix a\,(u)&0&0&0~\\  0&b\,(u)&c\,(u)&0~\\
0&c\,(u)&b\,(u)&0~\\ 0&0&0&a\,(u)~\endpmatrix
\tag 40$$
\par\noindent
where
$$a\,(u)=\sinh (u+\eta)\,,\quad b\,(u)=\sinh u\,,
\quad c\,(u)=\sinh\eta\,.
\tag 41$$

It is natural now from the above formulation of the problem
to call the RTL a $q$-deformation ($q=e^\eta$) of the ordinary Toda
lattice.

Separation of variables for the quantum RTL is taken in a complete
analogy to the separation for the ordinary Toda
lattice [10]. One considers the operator roots of the equation
$C_N\,(  u  )  =  0$. They are $N-1$ Hermitian
operators $y_j$ which give quantum analogs of the corresponding
real simple zeros in the classical case (36).
Let us introduce the operators $Y_j^\pm,\, j=1,\ldots,N-1\,$ by the rule:
$$ Y^-_j\,= A_N\,( u => y_j\, )\,,\qquad  Y^+_j\,= D_N\, ( u => y_j\,)\,.
\tag 42$$
These operators are defined by the values of the functions $A\,( u )$ and
$D\,( u )$ in $u  = y_j$.
The substitution of operator values for $u$ will be defined correctly if one
prescribes some rule for operator ordering. Here the ordering of
$y$'s to the left chosen.
We will call it substitution from the left
and denote $u  => y_j$.
The operators $y_j\,,~Y^\pm_j\,,~j = 1,\ldots,N-1$
have nice commutation relations
$$\bigl[ y_j\,, y_n\,\bigr] = \bigl[ Y^\pm_j\,, Y^\pm_n\,\bigr] = 0\,,\quad
Y^\pm_jY^\mp_j\,= 1\,,\quad
\bigl[y_j\,,Y^\pm_n\,\bigr]=\mp\,i\,\eta\,\delta_{jn}\,Y^\pm_n$$
which follow from the basic commutation relation (39) [10].

Let we consider the spectral problem
$$\tr T\,(u)\,\psi=\tau\,(  u )\,\psi$$
then substitute $u  => y_j$ (from the left) and use the definition
of $Y^\pm_j$.
The resulting set of equations has the form
$$ \tau\,( y_j)\,\psi\,( y_j\,) =
(\Delta_-\,Y^-_j + \Delta_+\,Y^+_j\,)\,\psi\,( y_j\, )
$$
where $\Delta_\pm\,(u)$ give a factorization
of the quantum determinant of the monodromy matrix [10]
$\Delta(u)= \qdet T\,( u )=\Delta_+\,\Delta_-=1.$ Let
$\Delta_-\,=i^{-N}\,,~\Delta_+\,=i^N$.
In $y$-representation the operators $Y^\pm_j$ are acting as shift operators
and the separation equations take the form
$$ \tau\,( y_j)\,\psi\,( y_j)
=\, i^N\,\psi\,(\,y_j\,+\,i\,\eta\,)\,+
\,i^{-N}\,\psi\,(\,y_j-\,i\,\eta\,)\tag 43$$
while the full eigenfunction can be represented in the factorised form
(quantum separation of variables)
$$\psi(y_1\,,\ldots,y_{N-1}\,)=\prod\limits_{k=1}^{N-1}\,\psi(y_k).$$
The separation equations above are $N-1$ one-dimensional finite-difference
multiparameter spectral problems.

We can rewrite equations (43) as
$$\aligned
\sum^N_{k=0}\,\Bigl(\,\,I_{N-2k}\,\exp\bigl(\,(N-2k)\,y_j\,\bigr)
\,\Bigr)\,\psi\,(y_j\,)\,=\qquad\qquad\\
\qquad\quad =\, i^N\,\psi\,(\,y_j\,+\,i\,\eta\,)\,+\,i^{-N}\,\psi\,
(\,y_j\,-\,i\,\eta\,)\endaligned
\tag 44$$
where $I_{\pm k}$ are the eigenvalues of the quantum integrals of motion.
For ordinary Toda lattice these equations have the form
$$\sum^N_{k=1}\,\bigl(I_k\,y_j^k\,\bigr)\,\psi\,(y_j\,)=
i^N\,\psi\,(y_j\,+i\,\eta)\,+\,i^{-N}\,\psi\,(y_j\,-i\,\eta)\,.
$$

As for a solution of the above separation equations, there are
more questions till now than answers, although in the non-relativistic
case one can use the Gutzwiller algorithm.
The RTL case provides even more difficult problem,
for example, for $N=2$  the separation equation
$$\bigl(\cosh p +E\,\bigr)\,\psi=\sinh x\,\psi\,$$
becomes the famous Harper equation [5].

\head
5. Monodromy matrices of the quantum RTL\\
for general classical root systems
\endhead
\noindent
The monodromy matrices for such systems obey, as well as in the
non-relativistic case [7,11], the reflection equations
$$\aligned
R\,(u-v )\,\overnumber{U_-}{1}\,( u )\,& R\,( u +
v - \eta )\,\overnumber{U_-}{2}\,( v )=\\
 \overnumber{U_-}{2}\,(v)& R\,( u+ v- \eta)\,
\overnumber{U_-}{1}\,(u)\,R\,( u- v)\,,\\
R\,(-u + v )\,\overnumber{U^{t_1}_+}{1}\,( u )\,&
R\,(- u - v - \eta )\,\overnumber{U^{t_2}_+}{2}\,( v )=\\
\overnumber{U^{t_2}_+}{2}\,(v)& R\,(- u - v - \eta)\,
\overnumber{U^{t_1}_+}{1}\,( v)\,R\,(- u+ v)\,,\endaligned
\tag 45$$
where $\,\overnumber{U_\pm}{1}=U_\pm\otimes I$,\;
$\,\overnumber{U_\pm}{2}=I\otimes U_\pm$ and
$t_{1,2}$ are matrix transpositions
in the first and second spaces, respectively.

We have the following rules for constructing the monodromy matrices
$\th(u)$ here [11]:
$$\aligned
U_-\,(u)&=T_-\,(u)\,K_-\,(u-\dfrac{i\eta}2\,)\,T_-^{-1}\,(-u)\,,\\
U^t_+\,(u)&=T_+^t\,(u)\,K_+^t\,(u+\dfrac{i\eta}2\,)\,
\bigl(T_+^{-1}\,(-u)\,\bigr)^t,\\
\th(u)&= U_+\,(u)\,U_-\,(u),\quad\text{for the closed lattices (interval)}\\
\th(u)&= U_-\,(u),\quad\text{for the open lattices (semi-axis)}
\endaligned
\tag 46$$
where the $T_\pm\,(u)$ are matrices obeying the  equation (39) and
the $K_\pm$ are simple solutions of the reflection equations (45).
The generating function of the integrals of motion is [7,11]
$$\aligned
t\,(u)&=\tr\th\,(u)=\tr\,U_+\,U_- \\
&=\tr\,K_+\,(u+\frac{i\eta}{2}\,)T\,(u)\,K_-\,(u-\frac{i\eta}{2}\,)\,
T^{-1}\,(-u)\,,\\
T\,(u)&=\prod\limits_{j=1}^M L_j\,(u)\,.
\endaligned $$
For the root systems $B_N$ and $C_N$, $M =  N$
and matrices
$K_\pm$  are scalar solutions of the reflection equations (45);
for the root system $D_N$,
$M = N-2$ and matrices  $K_\pm$  depend on the dynamical variables.

We list below the matrices $K_\pm$  and corresponding
Hamiltonians for the quantum generalized RTL's
(we give also the corresponding $K$-matrices for
the non-relativistic Toda lattices):
\vskip0.5 true cm\par\noindent
$B_N$ and $C_N$ series:
\vskip0.5 true cm\par\noindent
for RTL's
$$\aligned
K_-\,=\pmatrix
\frac{\alpha_1}{\sinh u}+\frac{\beta_1}{\cosh u} & -\gamma_1\\
\gamma_1 & \frac{\alpha_1}{\sinh u}-\frac{\beta_1}{\cosh u}
\endpmatrix\,,\\
K_+\,=\pmatrix
\frac{\alpha_N}{\sinh u}-\frac{\beta_N}{\cosh u} & \gamma_N\\
-\gamma_N & \frac{\alpha_N}{\sinh u}+\frac{\beta_N}{\cosh u}
\endpmatrix\,,
\endaligned
\tag 47$$
for non-relativistic Toda lattices
$$K_-\,=\pmatrix \alpha_1 & -u\\  \beta_1\,
u & \alpha_1\endpmatrix\,,\qquad
K_+\,=\pmatrix \alpha_N & -\beta_N\,u\\
-u & \alpha_N\endpmatrix \,.$$
The Hamiltonian of the corresponding RTL's is
$$H=H_0\,+J_1\,+J_N\,, \tag 48$$
where we use the notations
$$H_0\,=\sum\limits_{j=1}^N\,2\cosh (p_j\,)+2\,\sum\limits_{j=1}^{N-1}\,
\left[\exp(q_{j+1}\,-q_j\,)\cdot\cosh(\dfrac{p_{j+1}\,+p_j}{2})\right]\,,
\tag 49$$
$$\aligned
J_1\,&=
2\exp( q_1\,)\,\left[\alpha_1\cosh(\dfrac{p_1}{2})
+\beta_1\sinh(\dfrac{p_1}{2})\right]+\gamma_1\,\exp(2q_1\,)\,,\\
J_N\,&=
2\exp( q_N\,)\,\left[\alpha_N\cosh(\dfrac{p_N}{2})
+\beta_N\sinh(\dfrac{p_N}{2})\right]+\gamma_N\,\exp(-2q_N\,)\,.\endaligned
\tag 50$$
We have here RTL's associated to the following root systems:
$C_N\quad(\alpha_1\,=\alpha_N\,=\beta_1\,= 0)$,
$B_N\quad(\alpha_1\,=\beta_N\,=\beta_1\,= 0)$,
for the open lattices and
$C^{(1)}_N\quad(\alpha_1\,=\alpha_N\,= 0)$,
$A^{(2)}_{2N}\quad(\beta_1\,=\alpha_N\,= 0)$,
$D^{(2)}_{N+1}\quad(\beta_1\,=\beta_N\,= 0)$
for the closed lattices.

\vskip0.5 true cm\par\noindent
 $D_N$:
\vskip0.5 true cm\par\noindent
Denote
$\widehat{F_j}= \cosh\,(p_j\,+ q_j\,)$ and $\widehat{G_j}=\cosh\,(p_j\,-
q_j\,)$ then
$$\aligned
\widehat K_-\,&=\pmatrix \exp\,(u)\,\widehat{F}_1+\exp\,(- u)\,\widehat{G}_1&
\sinh^2\,q_1\,-\sinh^2\,( u )\\
       \sinh^2\,( u ) - \sinh^2\,p_1 & \exp\,( u )\,\widehat{F}_1 -
\exp\,(- u ) \widehat{G}_1\endpmatrix \,,\\
\vspace{20pt}
\widehat K_+\,&=\pmatrix \exp\,(u)\,\widehat{F}_N-\exp\,(- u)\,\widehat{G}_N&
\sinh^2\,( u ) - \sinh^2\,p_1 \\
       \sinh^2\,q_1\,-\sinh^2\,( u ) & \exp\,( u )\,\widehat{F}_N +
\exp\,(- u ) \widehat{G}_N\endpmatrix \,.\endaligned
\tag51$$
For ordinary Toda lattices
$$\aligned
\widehat K_-\,&=\pmatrix u\cosh q_1 - p_1\,\sinh q_1& \sinh^2\,q_1\,\\
                     u^2 - p^2_1 & u\cosh q_1 + p_1\,\sinh q_1 \endpmatrix
\,,\\
\vspace{20pt}
\widehat K_+\,&=\pmatrix u\cosh q_N + p_N\,\sinh q_N&u^2 - p^2_N \\
                 \sinh^2\,q_N&u\cosh q_N - p_N\,\sinh q_N \endpmatrix \,.
\endaligned $$
The Hamiltonian now is
$H=H_0\,+\tilde{J}_1\,+\tilde{J}_N$
where $H_0$ is given by (49) and
$$\aligned
\tilde{J}_1\,&=2\,\exp( q_1+q_2\,)\,
\cosh(\dfrac{p_1+p_2}{2})+\exp(2q_2\,)\,,\\
\tilde{J}_N\,&=2\,\exp(- q_N-q_{N-1}\,)\,
\cosh(\dfrac{p_N+p_{N-1}}{2})+\exp(-2q_{N-1}\,)\,,\endaligned
\tag 52$$
We have here RTL's associated to the following root systems:
$D^{(1)}_N\,,$ $B^{(1)}_N\,$  $(\beta_N\,= 0)$ and
$A^{(2)}_{2N-1}$ $(\alpha_N= 0)$.

\head 6. Acknowledgments
\endhead
The authors are grateful to I.V.~Komarov and E.K.~Sklyanin
for valuable discussions.
AVT acknowledges the support from the Netherlands Organisation
for Scientific Research
(NWO) during one-week stay at the
Universiteit van Amsterdam. VBK was supported by NWO under the Project
\# 611--306--540.

\enddocument